\documentclass{aa}  
\usepackage{graphicx}
\usepackage{xcolor}
\usepackage{txfonts}

\begin{document}

   \title{\texttt{Nemesis}: A Multi-Scale, Multi-Physics Algorithm for Astrophysics}

   \author{
        E. Hochart\inst{1} \and
        S. Portegies Zwart\inst{1}
    }
   \institute{
             Leiden Observatory, University of Leiden, 
             Niels Bohrweg 2, 2333 CA Leiden\\
             \email{hochart@mail.strw.leidenuniv.nl}
    }

   \date{Received 11/11/2025; accepted 10/03/2026}

  \abstract
   {Astronomical environments are governed by a complex interplay of physical processes, including gravitational dynamics, radiative transfer, stellar evolution, chemistry and hydrodynamics. These processes span a vast range of spatial scales, from short-range interactions where intra-particle distances are vital, to cosmological scales. Both these characteristics make astrophysics a multi-scale, multi-physics discipline. This multi-physics, multi-scale nature of astronomy introduces complications to numerical methods where round-off errors and a wide range of temporal timescales on which processes act can influence the reliability of results and the efficiency of algorithms. }
   {In this work, an updated version of the multi-scale, multi-physics algorithm, \texttt{Nemesis}, which makes use of the Astrophysical Multipurpose Software Environment (\texttt{AMUSE}). The algorithm is formally introduced and validated.}
   {A suite of simulations is run to assess its performance in simulating star clusters containing planetary systems, its ability to capture the von Zeipel–Lidov–Kozai effect, and its computational scalability.}
   {\texttt{Nemesis} is found to yield indistinguishable results in both the global and local scales when compared with the direct $N$-body code \texttt{Ph4}. The same conclusion is found when analysing its ability to capture the von Zeipel-Lidov-Kozai effect. When analysing its computational performance, the wall-clock time scales roughly as $t_{\rm sim}\propto1/\sqrt{\delta t_{\rm nem}}$ where $\delta t_{\rm nem}$ represents the time synchronisation between the global and local scales. When changing the number of planetary systems, the wall-clock time remains unchanged as long as the number of available cores exceeds the number of systems. Beyond this, it's found that at worst, the computational time increases linearly with the number of excess systems. }
   {The method introduced here can find its use in numerous domains of astronomy, thanks to its flexibility and modularity, from simulating protoplanetary disks in star clusters to binary black holes in the galactic centre.}


   \maketitle

\section{Introduction}
    In the early eighties, it was proposed that the Sun had a distant companion, `Nemesis'. Nemesis helped explain the apparent periodic epochs of extinction found in the fossil records \citep{1984Natur.308..713W, 1984Natur.308..715D}. Its existence has since been ruled out by infrared observations \citep{2011ApJS..197...19K} and additional data analysis showing that the mass extinction events are, in fact, not periodic \citep{2011MNRAS.416.1163B}. Nevertheless, the notion that such a distant body can be perceived as a companion highlights the multi-scale nature of astronomy. Compared to the $88$ days Mercury takes to orbit the Sun, Nemesis would take $\sim10^{10}$ days \citep{1984Natur.308..713W}, roughly the orbital period of the Sun around the galactic centre. 
    
    The multi-scale nature of astronomy complicates matters in computational astrophysics. A wide range in mass and length (i.e stars orbiting a supermassive black hole or clusters orbiting the galaxy) invokes numerical round-off errors \citep{2015ComAC...2....2B}. These numerical errors may grow exponentially in chaotic $N$-body systems \citep{1891BuAsI...8...12P,1892mnmc.book.....P, 1986LNP...267..212D}, resulting in diverging evolutionary histories. The matter is further complicated since a wide range in length can also translate to a wide range in characteristic time for dynamical systems. For instance, while a planet may orbit its host on the order of months or even days, a young stellar cluster will have crossing times of $\sim1$ Myr and may dissolve in the galactic tidal field on the order of $100$ Myr \citep{2010ARA&A..48..431P}. This multi-scale nature in dynamical systems has instigated the production of various $N$-body integration strategies over the decades, with each implementing their own schemes and approximations depending on the problem at hand (see, e.g. \citet{2003IAUS..208...81H} for a review):
    \begin{itemize}
        \item \textbf{Statistical Methods} \citep{1971ApJ...164..399S, 1971Ap&SS..14..151H, 1979ApJ...234.1036C, 1998MNRAS.298.1239G, 2005MNRAS.364..948S}: Statistical methods like the Monte Carlo, the anisotropic gaseous models, or Fokker-Planck methods are well suited for integrating systems satisfying certain symmetries, such as globular clusters. However, their statistical approach incurs poor solutions when resolving tight encounters or systems where spherical symmetry is no longer applicable.
        
        \item \textbf{Tree Codes} \citep{1986Natur.324..446B}: Tree codes compute pairwise gravitational forces between nearby particles, but approximate those with larger separation. This method is built for large $N$ problems, namely galactic and cosmological simulations, scaling as $\mathcal{O}(N\log N)$ with $N$ being the number of particles simulated. However, approximating long-range forces makes them ill-suited to resolving collisional systems, where energy conservation is required to be low to ensure interactions are accurately resolved.
        
        \item \textbf{Direct $N$-body codes}: These codes compute all pairwise forces. Their $\mathcal{O}(N^2)$ scaling make them poorer performers at large $N$ relative to the former two families mentioned, but techniques such as regularisation \citep{Stiefel1965, 1974CeMec..10..185A}, individual time steps or the Ahmad-Cohen scheme \citep{1973JCoPh..12..389A, 1992PASJ...44..141M} mitigate bottlenecks making them optimised for cluster dynamics. These same approximations affect the smallest scales and do not yield time-symmetric solutions. Following \citet{1918NGotM......235N}, without time-symmetry, a drift in energy errors is introduced, artificially altering a system's dynamics over secular timescales.
        
        \item \textbf{Symplectic codes} \citep{1991AJ....102.1528W}: Symplectic codes provide time-symmetric solutions thanks to their equations directly mapping from the system's Hamiltonian. As a result, symplectic codes are ideal for simulating planetary systems on long time scales. Nevertheless, their time-symmetric nature requires all particles to share the same time step. Constrained by the tightest orbit, they perform poorly beyond moderate $N$ systems. 
    \end{itemize}
    Hybrid codes attempt to stitch together different families to fit their problem best. For instance, \texttt{LonelyPlanets} \citep{2017MNRAS.470.4337C} adopts a two-step process to integrate planetary systems in star clusters, using both a symplectic code for its planetary orbits and a direct $N$-body code for the cluster evolution. Meanwhile, codes such as \texttt{P$^{3}$T} \citep{2015ComAC...2....6I}, \texttt{PeTar} \citep{2020MNRAS.497..536W} and \texttt{Ketju} \citep{2017ApJ...840...53R, 2023MNRAS.524.4062M} couple together tree code-like schemes with direct $N$-body integration to allow for optimised simulations of galactic environments, with special attention to the galactic core.

    The multi-physics discipline of astronomy adds another layer of complexity. For brevity, consider only a star cluster whereby not all the gas within a molecular cloud contributes to star formation. Observations of the solar neighbourhood indicate that $10\%\sim35\%$ of gas from a molecular cloud converts to stars \citep{2003ARA&A..41...57L, 2010ApJ...724..687L}, the rest remains embedded in the cluster, deepening the potential well and thereby affecting stellar dynamics. Once the first supernova occurs, this gas is expelled, and the resulting impulsive mass loss alters the cluster potential, driving rapid expansion \citep{1980ApJ...235..986H} and reshaping the gravitational potential, both of which influence the stellar dynamics. The influence of stellar evolution on dynamics is also apparent during a star's lifetime, whereby mass loss via stellar winds heats the cluster, causing both an expansion of the system and a reduction in the gravitational strength between stars. Beyond dynamics, stellar evolution also regulates the radiation field, which, for instance, controls the evaporation rate of protoplanetary disks and subsequently affects the growth of planets (see e.g. \citep{2022MNRAS.515.4287W, 2023MNRAS.520.5331W, 2024A&A...689A.338H}). These simple examples illustrate the extent to which gravitational, hydrodynamical, and radiative processes are intertwined in astronomy. Ignoring feedback between frameworks in astronomy risks oversimplifying problems, and while challenging to implement numerically, the inclusion of multi-physics in numerical models provides both more faithful and more predictive results.

    In this paper, we formally present a hybrid technique capable of handling multi-physics questions, \texttt{Nemesis}\footnote{https://github.com/spzwart/nemesis}. This technique was previously used to integrate planetary systems in star clusters \citep{2019A&A...624A.120V}; here, a more thorough description and validation of the recently modified and optimised version is provided in detail. Results show that the core approximation on which \texttt{Nemesis} is built, that is, decoupling widely separated systems from each other and solving them in parallel, yields negligible loss in physics at the reward of massive computational benefits. The flexibility this hybrid code provides to the user allows researchers to tackle multi-scale and multi-physics problems.

    \section{Theory of \texttt{Nemesis}}\label{Sec:Theory}
        \texttt{Nemesis} is developed within the \texttt{AMUSE} framework \citep{2009NewA...14..369P, 2013A&A...557A..84P, 2018araa.book.....P}. Currently, \texttt{Nemesis} is an external module and is not yet part of the \texttt{AMUSE} distribution officially. We intend to incorporate it into the main library in the future.
        
        \texttt{AMUSE} contains codes that can model different physical domains, such as stellar evolution, gravitational dynamics, astrochemistry, hydrodynamics, and radiative transfer. Within each domain, users can choose from multiple solvers, offering considerable freedom in tailoring simulations to specific scientific goals. A full list of available codes within \texttt{AMUSE} is provided in Appendix B of \citet{2018araa.book.....P}.
        
        This flexibility also becomes a perk in \texttt{Nemesis}, since it is designed to handle multi-physics problems with modularity. This is achieved via channels \citep{2014RSPTA.37230385V}. Channels communicate data between codes, allowing processes within one physical domain to influence another. \texttt{AMUSE} manages communication internally using the Message Passing Interface (MPI) \citep{GROPP1996789}, which helps minimise overhead during data transfer. Although \texttt{Nemesis} can be general purpose, its original motivation was to solve planetary systems in star clusters. The following discussion mostly considers this context.
    
    \subsection{Dealing with Scales}
        As mentioned in the introduction, the wide range of timescales present in star clusters containing planetary systems causes problems in computational astronomy. Although direct $N$-body algorithms can handle moderate-to-large $N$ systems, their various approximations introduce an energy drift, leading to unreliable solutions over secular timescales. This is especially problematic in environments where subsystems exist (e.g., planetary systems or binary stars) since these tighter systems evolve much faster than the larger-scale environment they inhabit. Symplectic codes enable accurate results over secular timescales; however, their shared time step scheme makes them inefficient for large $N$, as the tightest orbit halts all progress. 

        This clear separation between the characteristic evolutionary times of subsystems and the global environment allows one to decouple the problem and integrate each component in isolation, synchronising their states at regular intervals. As long as the synchronisation interval is short enough to capture essential interactions, the approximation holds. This separation motivates a hierarchical treatment of the environment, where the global and local dynamics are evolved independently. To describe this framework, the following terminology is introduced:
        \begin{itemize}
            \item \textbf{Children}: In \texttt{Nemesis} \textit{children} represent subsystems. That is, the set of particles lying within some distance of one another (see section \ref{Sec:HandlingChild}). A \textit{child system} can therefore correspond to, for example, a planetary system, a binary, or a triple. Within each child system, the individual particles (e.g., planets or stars) are referred to as \textit{child particles}.
            \item \textbf{Child code}: The \textit{child code} is the dedicated integrator assigned to a child system. Since each child system is integrated with its own code, users can ensure high energy conservation (e.g., by choosing a symplectic integrator). Additionally, this makes the scheme naturally parallelisable since child codes can be distributed across CPU cores, reducing run time.
            \item \textbf{Parent}: \textit{Parents} are the set of particles tracked in the global environment. A parent particle can represent either an isolated object (e.g., a single star or an interstellar object) or be a centre-of-mass (CoM) proxy of a child system.
            \item \textbf{Parent code}: The \textit{parent code} is the integrator used to track the global evolution of the set of parent particles.
        \end{itemize}
        Representing a child system with a CoM proxy ensures that tight orbits are removed from the parent code, bypassing the poor resolution of tight systems within a direct $N$-body integration while still preserving the optimised performance of the integrator for moderate-to-large $N$ systems.

        Over time, the number of children, $N_{\rm chd}$, may vary. During the parent integration, close encounters form a new child system. Conversely, a child system dissolves if any of its members become sufficiently separated (i.e., an ionising event between a binary and an intruder occurs). If $N_{\rm chd}$ exceeds the number of cores available, excess children wait in a queue until a CPU becomes available. CPUs are released the moment a child code finishes integrating, with priority given to the most computationally expensive jobs.
            
    \subsection{Synchronising Scales}\label{Sec:Synchronise}
        Since children are isolated from the global environment, neither scale communicates its dynamical history to the other. To ensure feedback across scales, \texttt{Nemesis} applies gravitational correction kicks between parents and children at regular time intervals, referred to as the bridge time step, $\delta t_{\rm nem}$. These kicks compensate for the fact that, in the parent code, children are represented by a CoM proxy. While this approximation reduces computational cost, it alters the underlying potential during the parent integration step. Correction kicks restore missing forces in two ways:
        \begin{itemize}
            \item \textbf{Correction on parents}: Each parent receives a correction equal to the difference between the gravitational force exerted by all child particles within a system, minus that of its representative parent. Formally,
        \begin{equation}
            F_{{\rm corr.\, par},\, i}=\sum_{j=0,j\neq i}^{N_{\rm chd}}\left(\sum_{k=0}^{n_{\rm chd}}F_{k} - F_{{\rm par},\, j}\right),
        \end{equation}
        where $F_{{\rm corr.\, par},\, i}$ is the correction force applied to parent $i$, $F_k$ are forces exerted by child particles contained within system $j$, and $F_{{\rm par}, j}$ is the force from the parent representative of child system $j$ (i.e the CoM representative integrated in the parent code). This correction is summed over all children, $N_{\rm chd}$. The condition $j \neq i$ ensures that a parent does not consider its own child system, as their dynamics are already accounted for in its phase-space representation.

        For a virialised Plummer sphere of density $\rho\sim10^{2}$ M$_\odot$ pc$^{-3}$, when $10/200$ of the stars host planetary systems, applying these corrections over bridge time steps of $500$ yr results in a kick of $\delta v\sim10^{-17}$ km s$^{-1}$. Increasing the density to $10^{3}$ M$_\odot$ pc$^{-3}$ and the number of planetary systems to $100/2000$, the corrections become $\delta v\sim10^{-13}$ km s$^{-1}$. 
        
        While the corrections are minor, they accumulate in time. Moreover, as mentioned earlier, large $N$ systems are inherently chaotic  \citep{1891BuAsI...8...12P,1892mnmc.book.....P, 1986LNP...267..212D}. As such, inclusion of these corrections may translate into subtle yet distinct dynamical effects.
        \end{itemize}
        \begin{itemize}
            \item \textbf{Correction on children:} Each child particle receives a correction equal to the difference between the total force from all external parents and the force from its own parent. Formally,
            \begin{equation}
                F_{{\rm corr.\, chd},\, j}=\sum_{j=0}^{N_{\rm chd}}\left(\sum_{k=0,k\neq j}^{N_{\rm par}}F_{k} - F_{{\rm par},\, j}\right),
            \end{equation}
            where $F_{{\rm corr.\, chd},\, j}$ is the gravitational correction applied onto all child particles within system $j$, $F_k$ is the force from all external parents, and $F_{{\rm par}, j}$ is the parent of system $j$.
            
            As before, to give context with numbers, a virialised Plummer sphere of density $100$ M$_\odot$ pc$^{-3}$ having $10/200$ of the stars hosting planetary systems, the corrections applied to the planets over the same $500$ yr bridge time steps results in velocity shifts of $\delta v\sim10^{-10}-10^{-7}$ km s$^{-1}$. The large spread is due to the sensitivity of the local environment, with gravity having an inverse-squared law with spatial separations. When $\rho\sim10^{3}$ M$_\odot$ pc$^{-3}$, this ranges between $\delta v\sim10^{-8}-10^{-5}$ km s$^{-1}$.
        \end{itemize}
        Correction kicks are applied every bridge time step via a second-order Verlet kick-drift-kick method \citep{1967PhRv..159...98V, 2007PASJ...59.1095F}. In the limit $\delta t_{\rm nem}\xrightarrow[]{}0$ yr, \texttt{Nemesis} mimics a direct $2$nd-Order $N$-body integrator, although one with substantial overhead. In the opposite limit, $\delta t_{\rm nem}\xrightarrow[]{}\infty$ yr, computational time is minimised, but accuracy is lost.
        
        The only interaction not considered in \texttt{Nemesis} is between members of distinct children. Concretely, particles within one planetary system will not influence particles within another planetary system. These particles will only interact through their parents. Given this, the distance criteria for forming a child system should remain small relative to the interparticle distance. This will enhance the validity of the CoM approximation and minimise situations where several comparable-mass particles reside in the same child system, since this reduces the accuracy of the CoM approximation.

        Once a particle is no longer identified as a child, for example, when it is ejected and becomes a rogue planet, it becomes a parent, after which its dynamical evolution will be treated in the parent code, and it will feel the effects of all still existing children through correction kicks (see section \ref{Sec:Child_Dissolutions}).
        
        \subsection{Handling Parent Mergers}\label{Sec:HandlingChild}
            At times, two parents will be near enough that their internal dynamics require more accurate resolution. Such events are detected via \texttt{AMUSE}'s collisional detection stopping condition, which checks at every internal time step whether two parents are spatially separated by a distance less than the sum of their collisional radii. In \texttt{Nemesis}, a parent's collisional radius is defined as,
            \begin{equation}
                R_{\rm par} = A\left(\frac{M_{\rm par}}{{\mathrm M}_\odot}\right)^{1/3}\, {\rm  au}. \label{Eqn:Rpar}
            \end{equation}
            Here $A$ is a user-defined coefficient (default $A=100$) and $M_{\rm par}$ is the mass of the parent. For isolated parents (i.e., parents hosting no children, such as a rogue planet or single star), $M_{\rm par}$ is the particle's mass. If, instead, the parent hosts children, $M_{\rm par}$ is the total mass of the child system. By default, an upper limit of $R_{\rm par}=10^{3}$ au is applied. While test particles are massless and therefore have no radii, isolated test particles can still merge upon the approach of another (massive) parent particle. \texttt{Nemesis} handles mergers using two general schemes:
            \begin{itemize}
                \item \textbf{Case 1: At least one parent particle hosts children}\\
                This case is flagged when, for example, a planetary system has a close approach with an unbound star or two planetary systems encounter each other. In this situation, the correction kicks between participating child particles applied during the previous bridge time step are removed. A new code is spawned and assigned to the set of participating child particles which form the new child system. To minimise numerical error, child codes are integrated in the CoM frame of the child system.\\
                \item \textbf{Case 2: Neither parent particle hosts children}\\
                This case is flagged when, for example, two solitary stars undergo a close encounter. In this case, the event proceeds as in a standard $N$-body calculation, since no correction kicks were applied between the particles during the last bridge time step. Both parents now form a new child system and are assigned a dedicated child code to resolve its internal dynamics.
            \end{itemize}
            
            In both cases, the encountering parents are removed from the parent code and replaced by a new parent. This new parent has its mass equal to the sum of the two parents (by definition, the sum in mass of all constituents within the new child system) and phase-space coordinates equal to their CoM.
            
        \subsection{Handling Children Dissolutions}\label{Sec:Child_Dissolutions}
            Just as parent particles can approach one another and form a new child system, child particles can also move apart. This can happen, for instance, with high velocity stars in the galactic centre or ejected debris in protoplanetary disks, etc... 
            
            In \texttt{Nemesis}, membership within a child system is determined using a friends-of-friends criterion with linking length $\alpha R_{\rm par}$ and $\alpha$ being a tunable parameter set to two by default. 
            
            At each bridge time step, particles are grouped into connected components by linking all pairs with separations $r_{ij} \leq \alpha R_{\rm par}$. The set of particles attributed to a connected component forms a child system and will have its internal dynamics resolved using a dedicated child code. If, during the evolution, a child system fragments into multiple connected components, the system is flagged as dissolved and is handled as follows:
            \begin{itemize}
                \item \textbf{Case 1: Child particle still has nearby neighbours} \\
                If the fragment contains two or more particles linked by separations $r_{ij}\leq \alpha R_{\rm par}$, a new child system containing those neighbours are created. This child system is assigned its own child code, and a new parent is introduced in the parent code. As before, this parent has its phase-space coordinates equal to the CoM of the child system and mass equal to the total child system's mass. \\

                \item \textbf{Case 2: Child particle is isolated}\\
                If the fragment consists of only one particle. That is, the particle has no neighbours within $r_{ij}\leq\alpha  R_{\rm par}$, it is no longer part of any child system. The particle is isolated and only followed in the parent code. Its attributes match those of the escaping particle, though shifted to the cluster frame of reference. Additionally, its radius no longer reflects its physical size, but instead follows equation \ref{Eqn:Rpar} to account for any future merging events.
            \end{itemize}
            Finally, child systems are only assigned child codes if they contain at least one massive particle. Groups composed solely of test particles never form a child system. Instead, each of these particles is added directly to the parent code as an isolated particle (an isolated parent), regardless of proximity.

    \subsection{Code Coupling}
        Presently, \texttt{Nemesis} works by coupling gravity with stellar evolution. By default, the $4$th-order Hermite integrator \texttt{Ph4} \citep{2012ASPC..453..129M} is used as the parent code, although other codes are available and can replace it by changing two lines: the import line and code instantiation (see the \texttt{README.md} file on GitHub). The only requirement for the parent code is that it is capable of detecting collisions to allow for the merging between parents.
        
        The user is also free to choose the code for the children. For a non-relativistic two-body system, the Kepler solver \texttt{TwoBody} \citep{2012NewA...17..711P} is appropriate, but symplectic codes such as \texttt{Huayno} \citep{2012NewA...17..711P} and \texttt{Rebound} \citep{2012A&A...537A.128R} are also available. If, instead, the user wishes for an arbitrary precision solver capable of handling relativistic effects, \texttt{Brutus} \citep{2015ApJS..216...29B, 2021PhRvD.104h3020B} is also available. As mentioned in the opening of this section, a full list of available codes within \texttt{AMUSE} is provided in Appendix B of \citet{2018araa.book.....P}. The modularity of \texttt{AMUSE} also allows modelling of different physical regimes. For instance, coupling a hydrodynamical code to the child code can be managed using another \texttt{BRIDGE} \citep{2007PASJ...59.1095F} layer, and allows analysis of protoplanetary disks in star clusters. One can even extend this by considering radiative feedback from the cluster environment to account for the evaporation of the disks.
        
        By default, \texttt{Nemesis} embeds the environment within a Milky Way-like potential using the \texttt{BRIDGE} method. The potential follows \citet{2015ApJS..216...29B} and contains a spherical nucleus and bulge \citep{1990ApJ...356..359H}, a Miyamoto-Nagai disk \citep{1975PASJ...27..533M} and a Navarro-Frenk-White potential \citep{1997ApJ...490..493N} for the dark matter halo. A suite of other background potentials is available within the \texttt{AMUSE} library. This option can be toggled off using the input variable \texttt{gal\_field}.
    
\begin{figure*}
            \includegraphics[width=.9\textwidth]{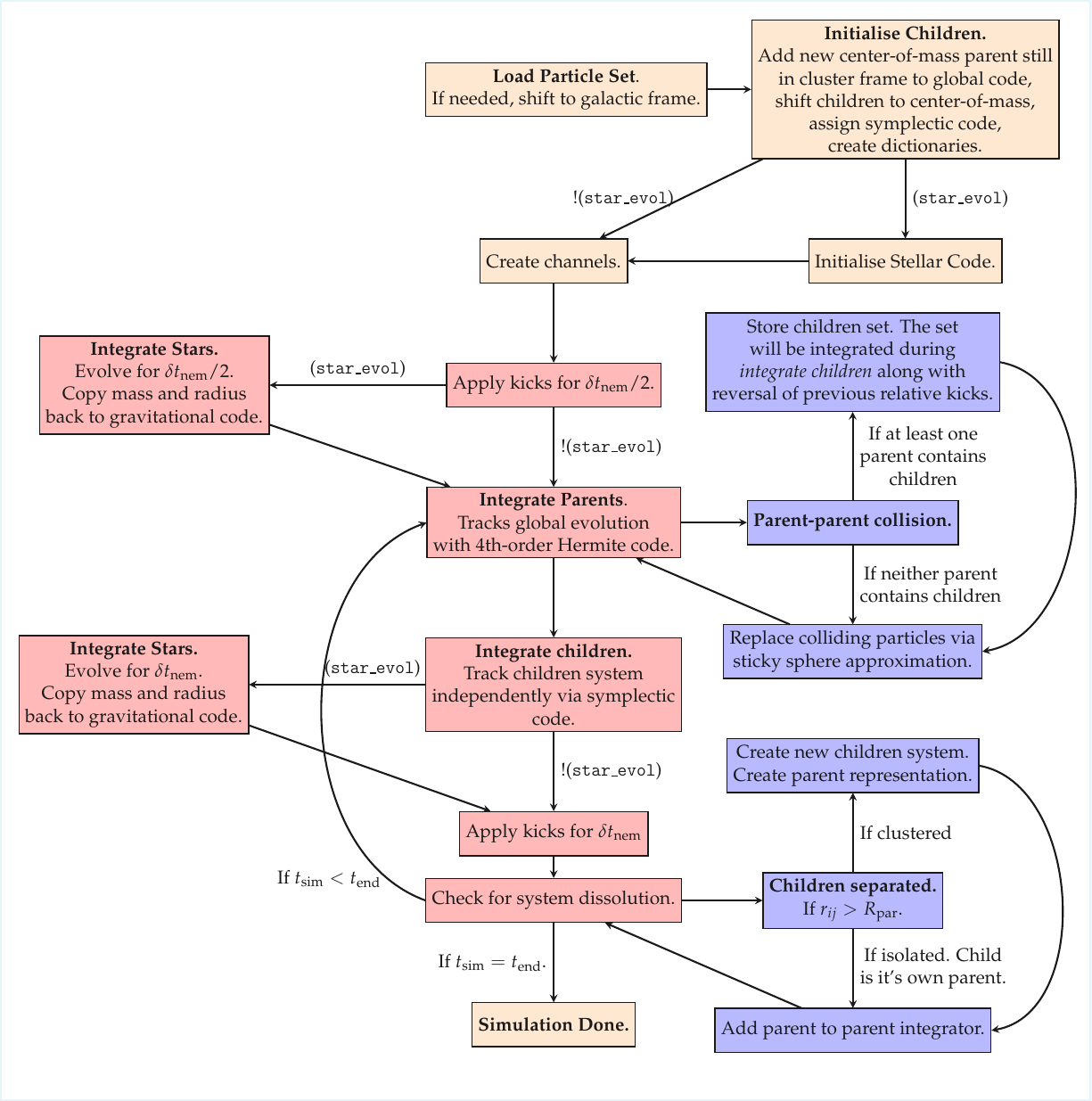}
            \caption{\texttt{Nemesis} workflow. \texttt{star\_evol} is a parameter in \texttt{Nemesis} toggling on or off stellar evolution. Communication between codes is internally handled by \texttt{AMUSE} via channels. Orange boxes represent the setup and termination of the simulation, red boxes indicate steps where work is offloaded to external codes, and blue boxes denote tasks handed back to Python for manipulation of \texttt{Nemesis}-structured data. A movie showing \texttt{Nemesis} in action is available online.}
            \label{Fig:NemesisWorkflow}
        \end{figure*}

        Stellar evolution defaults to using the parameterised code \texttt{SeBa} \citep{1996A&A...309..179P, 1998A&A...332..173P, 2001A&A...365..491N, 2012A&A...546A..70T}. \texttt{SeBa} can handle binary evolution and provides a quick way to resolve the general evolution of a star, although it doesn't provide its internal structure. A stopping condition is enabled to flag any star undergoing a supernova. If triggered, a natal kick is applied to the remnant star in focus. If one wishes to reproduce the rejuvenation and collision of stars better, one can adopt a Henyey solver (i.e \texttt{MESA}, \citet{2011ApJS..192....3P}) instead. Figure \ref{Fig:NemesisWorkflow} shows the general workflow of \texttt{Nemesis}. A video can also be accessed online.

\section{Validation and Tests}\label{Sec:Method}
    Throughout the paper, the $4$th-order Hermite integrator \texttt{Ph4} \citep{2012ASPC..453..129M} is used as the parent code (section \ref{Sec:Nemesis}). Children are integrated with \texttt{Huayno} unless otherwise stated. The parameter $\alpha$, which controls the linking length used to see if a child system has fragmented or not, is fixed to $\alpha = 2$. In all cases, the internal time step parameter is set to $\eta=0.1$. 
    
    The internal time step parameter linearly relates to the discretised time step with which the $N$-body integrator solves the equation of motion with the minimum inter-particle free-fall time scale, that is,
    \begin{equation}
        \delta t=\eta\,\,{\rm min}\left(\frac{|r_{ij}|}{|v_{ij}|},\, \sqrt{\frac{r_{ij}}{a_{ij}}}\right), \label{Eqn:Internaldt}
    \end{equation}
    where $\delta t$ is the discretised time step, and $r_{ij}$, $v_{ij}$, $a_{ij}$ the relative position, velocities and acceleration of particles $i$ and $j$ respectively, where $j\neq i$. Stellar evolution and galactic tides are turned off throughout the paper to facilitate comparisons between algorithms.

    To ensure consistency between runs, when testing \texttt{Nemesis}'s computational performance versus a direct $N$-body code (section \ref{Sec:Runs1}) and when changing different simulation parameters (section \ref{Sec:CompScaling_Methods}), the same computer cluster was used. This cluster comprises of $32$ Intel(R) Xeon(R) CPU E7-4820 processors clocked at 2.00 GHz.
    
    \subsection{Validation 1: Direct $N$-Body vs. \texttt{Nemesis}}\label{Sec:Runs1}
        The first set of simulations analyses the reliability of \texttt{Nemesis} (Section \ref{Sec:Nemesis}). To do so, the paper assumes that the direct $N$-body integrator (\texttt{Ph4}) provides ground truth results. That is, the more similarity between \texttt{Ph4} and \texttt{Nemesis}, the more reliable \texttt{Nemesis} is deemed to be. 
        
        Here, \texttt{Ph4} is used to integrate both the parent and children during \texttt{Nemesis} runs. The same integrator is used for the direct $N$-body run to facilitate comparison of results between algorithms and minimise computational load. A softening of $\epsilon=0.1$ au is applied for both schemes to further reduce computational load.

        The cluster consists of $N_*=2048$ stars, distributed in a virialised Plummer \citep{1911MNRAS..71..460P} sphere with half-mass radius $r_h=0.5$ pc. Stellar masses are drawn from a Kroupa mass distribution \citep{2001MNRAS.322..231K} ranging between $0.08-30$ M$_\odot$. Forty stars with masses between $0.5 - 2$ M$_\odot$ are randomly sampled to host a planetary system ($N_{\rm chd}=40$). The cluster is integrated until $0.1$ Myr, which is enough time for systematic errors to become apparent in the analysis. For \texttt{Nemesis}, a bridge time step of $\delta t_{\rm Nem}=500$ yr is adopted. 
        
        Planetary systems consist of one to six planets, whose mass and semi-major axis follow the oligarchic growth model \citep{2002ApJ...581..666K, 2015ApJ...807..157T}, and $250$ asteroids, represented as test-particles ($m_{\rm ast}=0$ kg). The inner disk edge corresponds to a $10$ yr orbit, while the outer-edge $R_{\rm out}=117 M_{\rm host}^{0.45}$ au \citep{2023MNRAS.520.5331W}. The asteroid disk has a Safronov-Toomre $Q$-parameter $Q=1$ \citep{1960AnAp...23..979S, 1972ApJ...178..623T}, and are distributed with surface density $\Sigma\propto r^{-3/2}$. Systems are randomly oriented along the ecliptic plane before any integration. 
        
        An additional run where \texttt{Nemesis} defaults back to integrating children with \texttt{Huayno} is also conducted to compare energy conservation. The internal time-step and softening parameter remain the same ($\eta=0.1$ and $\epsilon=0.1$ au), but the simulation runs until $1$ Myr.
        
    \subsection{Validation 2: von Zeipel-Lidov-Kozai Effect}      
        The second set of runs (section \ref{Sec:ZLK}) looks at the von Zeipel-Lidov-Kozai effect (ZLK, \citep{1910AN....183..345V, 1962AJ.....67..591K, 1962P&SS....9..719L}). The ZLK effect is a dynamical phenomenon in hierarchical triple systems, in which angular momentum is periodically exchanged between the inner and outer orbits when their mutual inclination exceeds a critical angle, $i_{\rm crit}=\arctan{\left(\sqrt{3/5}\right)}\approx39.2^\circ$. Because the component of angular momentum parallel to the total angular momentum vector of the system, $L_z=\sqrt{1-e^2}\cos{i}$, and semi-major axis (orbital energy) remains conserved, this interaction leads to coupled oscillations in the eccentricity and inclination of the inner orbit.
        
        The hierarchical triple is set up identically to that in Figure 3 of \citet{2013MNRAS.431.2155N}. Explicitly, the system contains an inner-binary comprised of particles with masses $M_1=1$ M$_\odot$ and $M_2=1$ M$_{\rm Jup}$, and orbiting with semi-major axis $a=6.0$ au and eccentricity $e=10^{-3}$. The outer particle has mass $M_3=40$ M$_{\rm Jup}$ and orbits the inner binary's CoM with $a_{\rm out}=100$ au and $e_{\rm out}=0.6$. The initial inclination of the outer binary is $i_{\rm out}=65^{\circ}$ relative to the inner's orbital plane.

        For each configuration, three tests are conducted. The first considers a regime with no parent-parent mergers, that is, $R_{\rm par}=0$ au. Another considers the regime where the child system never dissolves, that is, $R_{\rm par}=10^{3} {\rm \, au}>2a_{\rm out}$. The last considers a scenario with frequent parent-merging and child system dissolution events, at times even both occurring within the same bridge time step, $R_{\rm par}=10^{2} {\rm \, au}$. Unlike the former two configurations, which replicate results from a direct $N$-body code, the latter configuration scrutinises the CoM approximation used to represent a child system within the parent integrator. It also tests \texttt{Nemesis}'s capability to handle parent-parent merging events and child system dissolution. Systems are integrated for $10$ Myr with a bridge time step of $\delta t_{\rm nem}=500$ yr. For reference, this is half the orbital period of the outermost body.

        In reality, when large eccentricities are reached (corresponding to flips in inclinations), tidal effects should be considered, and collisions may occur. In the case here and assuming a solar radius object for our star-like particle, collisions will occur when the inner binary achieves $(1-e)<7.7\times10^{-4}$. Although \texttt{AMUSE} allows for collision detection and contains a symplectic integrator solving tides in its library (\texttt{TIDYMESS}, \citet{2023MNRAS.522.2885B}), this is omitted here to allow scrutiny on \texttt{Nemesis} over longer times.
        
    \subsection{Computational Scaling: \texttt{Nemesis} Scaling Relations}\label{Sec:CompScaling_Methods}
        The last set of runs investigates the computational scaling relations of \texttt{Nemesis}. The simulated cluster environment is set up similarly to that in section \ref{Sec:Runs1}. Planets follow the oligarchic model once more; however, the number is increased to eight per system, and asteroids are no longer present. Three cluster configurations are considered: A massive, dense cluster with $N_*=2048$ and $r_h=0.5$ pc, a low-populated, sparse cluster with $N_*=256$ and $r_h=0.5$ pc and, finally, a massive, sparse cluster with $N_*=2048$ and $r_h=1$ pc. The latter two have equal densities. In all cases, the cluster is sampled from a Plummer distribution and initially in virial equilibrium. Clusters are integrated until $t=0.1$ Myr. 
        
        When analysing the dependence of computational resources on $\delta t_{\rm Nem}$, the number of planetary systems is initially fixed to $N_{\rm chd}=20$ in all three configurations while $\delta t_{\rm nem}$ varies between $10^{2}-10^{3}$ yrs, in equal increments of $10^{2}$ yr. When testing how the number of children influences computational resources, only the massive, dense cluster is considered, and the bridge time step is fixed to $\delta t_{\rm Nem}=10^{3}$ yr. In this case, $N_{\rm chd}$ initially ranges between $20-60$, spaced out in equal increments of ten. Although the same cluster realisation is used for the initial condition in all runs for the latter test, by changing $N_{\rm chd}$, some runs have stars hosting planetary systems which otherwise do not in others.
        
    \section{Results and Discussion}\label{Sec:Results} 
        \begin{figure}
            \centering
            \includegraphics[width=\columnwidth]{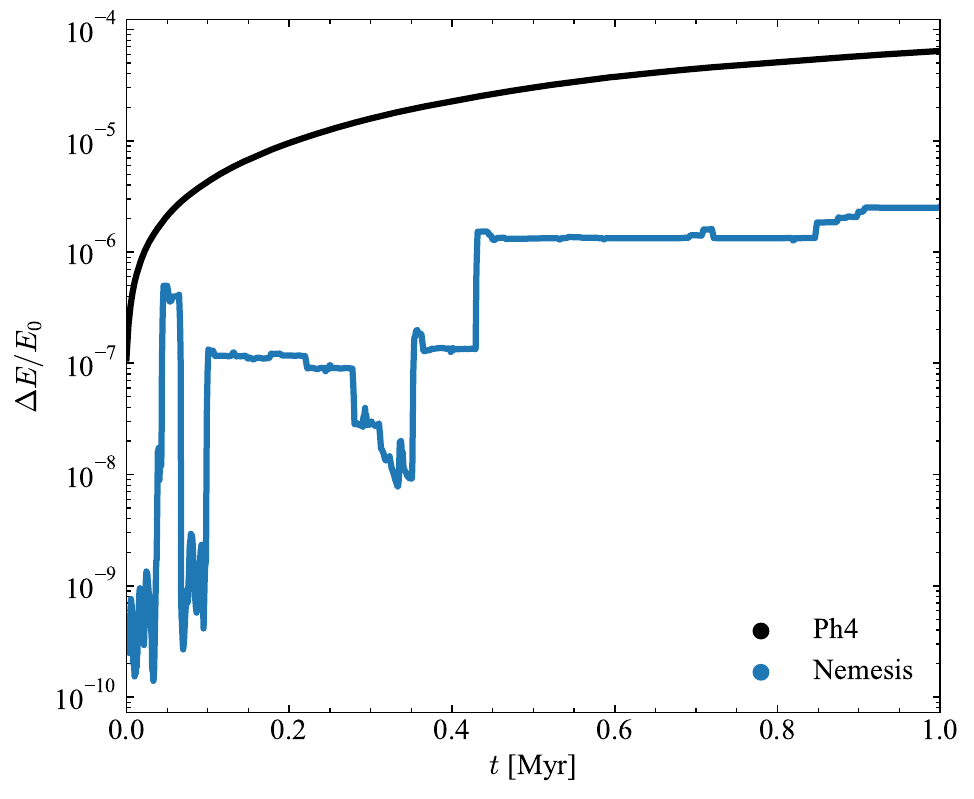}
            \caption{Evolution of the simulation's energy error in time.}
            \label{Fig:dE_Evol}
        \end{figure}
        \subsection{Direct $N$-Body vs. \texttt{Nemesis}}\label{Sec:DirectNbody}  
        The advantages provided by \texttt{Nemesis}'s flexibility in allowing for a symplectic algorithm for its set of children are evident in figure \ref{Fig:dE_Evol} since its relative energy error remains essentially bounded (see figure A.4 of \citet{2019A&A...624A.120V} for another example). Jumps, however, still appear due to the symplectic nature of \texttt{Huayno} hindering its ability to resolve close approaches. 
        
        In the purely direct $N$-body computation, \texttt{Ph4} exhibits a steady drift in energy error, which over longer timescales will yield unreliable results, especially for quickly evolving systems such as planetary systems. Note that $\eta=0.1$ was used for both runs, and energy errors can be reduced by adjusting this.
        
\begin{figure}
            \centering
            \includegraphics[width=\columnwidth]{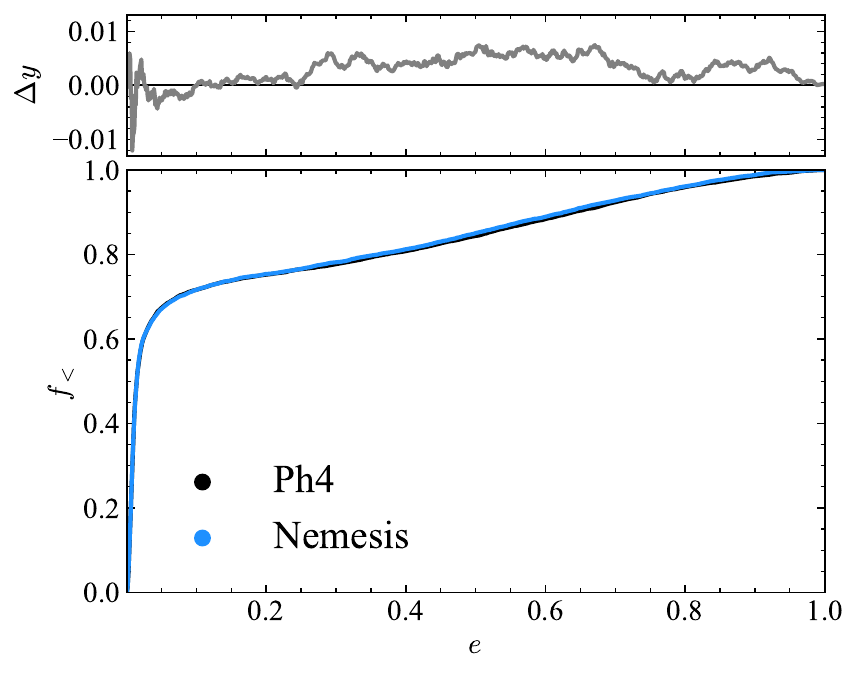}
            \caption{Bottom: Cumulative distribution function of asteroid eccentricities after $t_{\rm end}=0.1$ Myr. Top: Residuals in distributions between \texttt{Nemesis} and \texttt{Ph4}, computed as $\Delta y(e)=(y_{\rm \texttt{Nem}}(e)-y_{\rm \texttt{Ph4}}(e))$.}
            \label{Fig:CDF_Plots_ecc}
        \end{figure}

\begin{figure}
            \centering
            \includegraphics[width=\columnwidth]{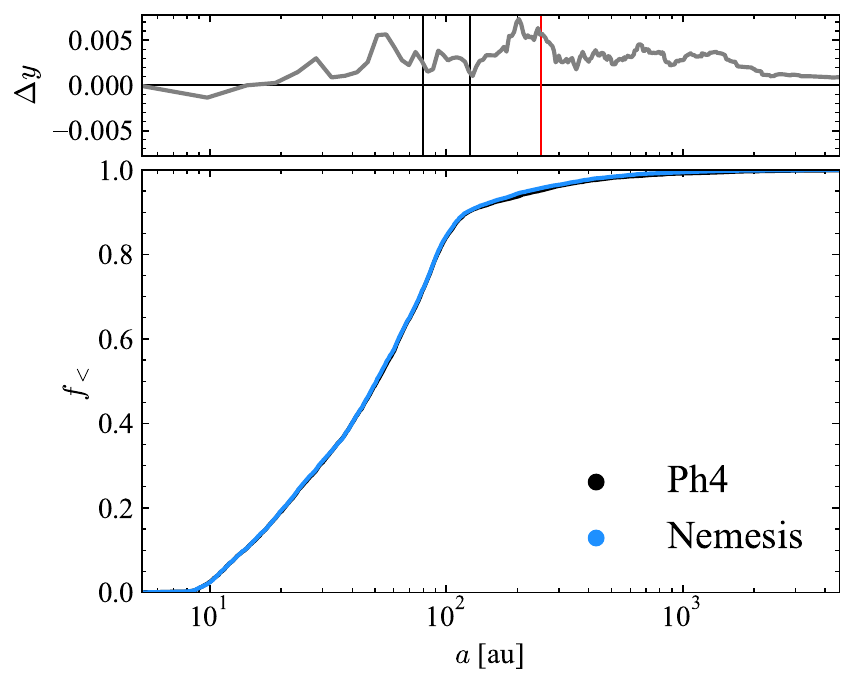}
            \caption{Same as figure \ref{Fig:CDF_Plots_ecc} but for asteroid semi-major axis. The two vertical black lines in the upper panel represent the range in parent radius initially considered ($0.5\leq M_{\rm par}$ [M$_\odot$]$\leq2.0$) when using equation \ref{Eqn:Rpar} and $A=100$. The red vertical line represents the linking length parameter, which flags children dissolution (section \ref{Sec:Child_Dissolutions}) for a $M_{\rm par}=2$ M$_\odot$ parent host.}
            \label{Fig:CDF_Plots_sma}
        \end{figure}

        For the remainder of this section, results correspond to the other set of simulations where \texttt{Nemesis} resorts to setting both the parent and child code as \texttt{Ph4}. The cumulative distribution in semi-major axis, $a$, and eccentricity, $e$, for all bound asteroids at the simulation's end is shown in figure \ref{Fig:CDF_Plots_ecc} and figure \ref{Fig:CDF_Plots_sma}. 
        
        A Cramer-von Misses \citep{cramer1928composition} test is done to see whether there are statistically significant differences in results. This test can compare the distribution spanning the total range of values, something lacking in the Kolmogorov-Smirnov test, which only probes the region with the largest deviation. A $p$-value of $p_{a}=0.98$ and $p_{e}=0.93$ is found for $a$ and $e$, respectively, implying that the distributions are indistinguishable.
        
        
        Given the importance of the tail-end distribution in an astronomical context since high-$e$ orbits can induce system instabilities, and/or result in mergers/ejection events, a two-sample \citet{Anderson01121954} test is also used. In the end, a $p$-value of $p_{a}=0.93$ and $p_{e}=0.50$ is found, allowing us to confidently reject the null-hypothesis that either samples are drawn from different distributions. The large-scale cluster properties also exhibit similar results with both the distribution in velocities and position of stars relative to one another having $p=1$.
        
        The similarity between distributions is also represented in the residuals shown on the upper panels of both plots. Residuals peak at $\max(\Delta y(e))\approx0.012$ and $\max(\Delta y(a))\approx7\times10^{-3}$ respectively. In other words, there is at most a $\lesssim 1.5\%$ difference in population occupancy at any point in orbital-parameter space between the two algorithms. For semi-major axis, \texttt{Nemesis} seems to underestimate the semi-major axis of particles occupying the $a\sim R_{\rm par}$ regime. 
        
        The residuals remain low and are likely attributed to statistical noise, a feature which naturally arises given the chaotic nature of many-body dynamical systems. Regardless, it is necessary to mention that \texttt{Nemesis} does perform worse when resolving the dynamical evolution of particles with semi-major axis $a \sim R_{\rm par}$, as indicated by the top panel of figure \ref{Fig:CDF_Plots_sma}. The origin of this discrepancy lies in the synchronisation between children and the global environment occurring at discrete time intervals (recall section \ref{Sec:Synchronise}).

        In \texttt{Nemesis}, correction kicks are only applied once every fixed bridge time-steps, $\delta t_{\rm nem}$. These bridge time steps are significantly larger than the internal integration time step $\delta t$ used by gravitational solvers such as \texttt{Ph4} (equation \ref{Eqn:Internaldt}). Indeed, while \texttt{Ph4} accounts for the gravitational influence of all particles at every internal time step, child particles in \texttt{Nemesis} experience the external potential only at the bridge time steps. In between these bridge time steps, child particles only interact with those part of the same child system.
        
        By having poorer resolution of external interactions and being more likely to compute the potential when fly-by events are already relatively far away, \texttt{Nemesis} underestimates the effective potential. Naturally, it is the child particles orbiting at $a\sim R_{\rm par}$ which are the most affected by this reduced resolution of the external potential.
        
        While decoupling children systems from the global environment allows for a computational speed up (cf. section \ref{Sec:Scaling}), this inevitably leads to a poorer representation of external perturbations. This is an inherent trade-off of the method. As mentioned in section \ref{Sec:Synchronise}, while setting $\delta t_{\rm nem}\xrightarrow[]{}0$ yr will give identical results to a direct $N$-body code, it will render the algorithm wildly inefficient.
        
        \begin{figure}
            \centering
            \includegraphics[width=.9\columnwidth]{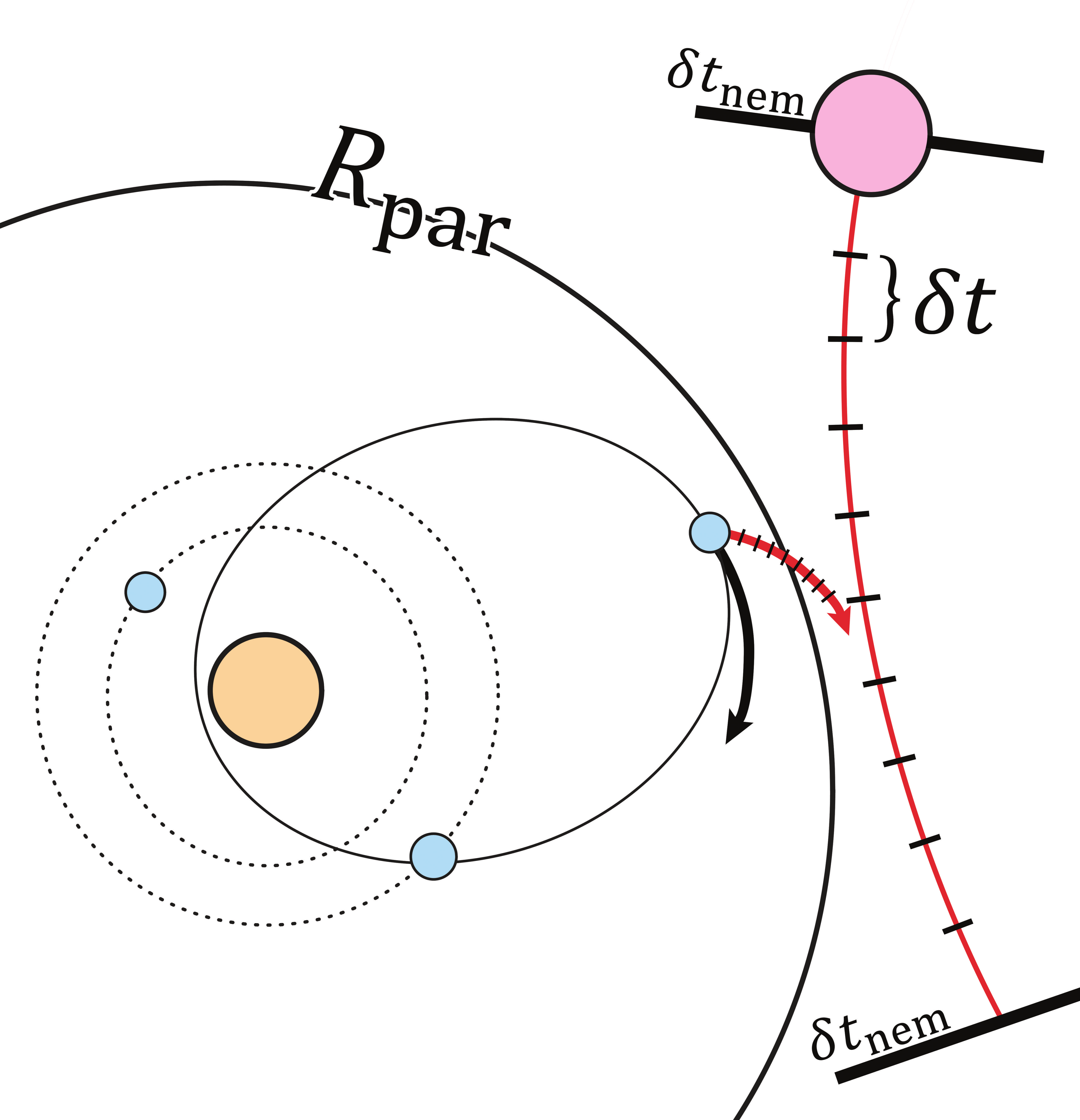}
            \caption{Diagram illustrating how children on wide orbits are not as well resolved in \texttt{Nemesis} since it manages external interactions via correction kicks applied every bridge time-step ($\delta t_{\rm nem}$). As a result, the outer planet (blue) is diverted along the black trajectory. Meanwhile, a direct $N$-body integrator computes interactions at every internal time-step ($\delta t$). With better resolution of the close encounter, the planet feels a stronger gravitational effect and is scattered outwards along the red path.}
            \label{Fig:DiagramResidual}
        \end{figure}
        To make the explanation more concrete, consider the diagram in Figure \ref{Fig:DiagramResidual}. Here, a planetary system composed of three planets (in blue) and a host star (in yellow) is experiencing a fly-by from an external star (in pink). Although the perturber approaches the system, it never reaches a separation $\delta r_{ij} \leq R_{{\rm par}, i} + R_{{\rm par},j}$, with $R_{{\rm par},j}$ being the child system’s parent radius. As a result, using \texttt{Nemesis}, no parent merger is triggered during the parent integration and the gravitational interaction of the fly-by is treated exclusively in the parent code between relevant parents, and during corrections kicks between the perturber and the planetary system.
        
        A direct $N$-body code can accurately resolve this encounter by integrating the equations of motion at each internal time step (represented schematically by tracks on the fly-by trajectory). In contrast, under the fixed bridge implementation of \texttt{Nemesis}, where each correction kick is computed at increments of $\delta t_{\rm nem}$, close interactions are likely to be under-resolved or missed. Consequently, although correction kicks are applied, the algorithm predominantly samples the external potential when the perturber is already relatively distant. This effect is only significant in moderate encounters. In the case that $\delta r_{ij} \leq R_{{\rm par}, i} + R_{{\rm par},j}$, the two systems will merge and all relevant particles will form a new children system where their internal dynamics will be resolved using a dedicated symplectic integrator (section \ref{Sec:HandlingChild}).
        
        In summary, due to \texttt{Nemesis}'s rigid bridge time step implementation to synchronise local and global scales, fly-by encounters not near enough to trigger parent merging events will be poorly resolved since the effective potential is likely to be computed only once the perturber is already far away. In turn, the effective external perturbation in \texttt{Nemesis} is weaker, and asteroids (or particles in general) are less likely to be scattered to large semi-major axes or high eccentricities. This is schematically illustrated in Figure \ref{Fig:DiagramResidual} with the diverging paths taken by the planet under \texttt{Nemesis} (black arrow) and under \texttt{Ph4} (red arrow). 
        
        While this has little impact on particles deep within the host potential, as reflected by the low residuals at small $a$ in figure \ref{Fig:CDF_Plots_sma}, it becomes important near the boundary of children systems. This region represents where particles are the least tightly bound to their host, and thus the most fragile to the external environment. Implementing an adaptive bridge time-step scheme in \texttt{Nemesis} is one of the improvements currently under consideration, with preliminary results using reinforcement learning being encouraging \citep{2025CNSNS.14508723S}.

    \subsection{von Zeipel-Lidov-Kozai Effect}\label{Sec:ZLK} 
        \begin{figure}
            \centering
            \includegraphics[width=\columnwidth]{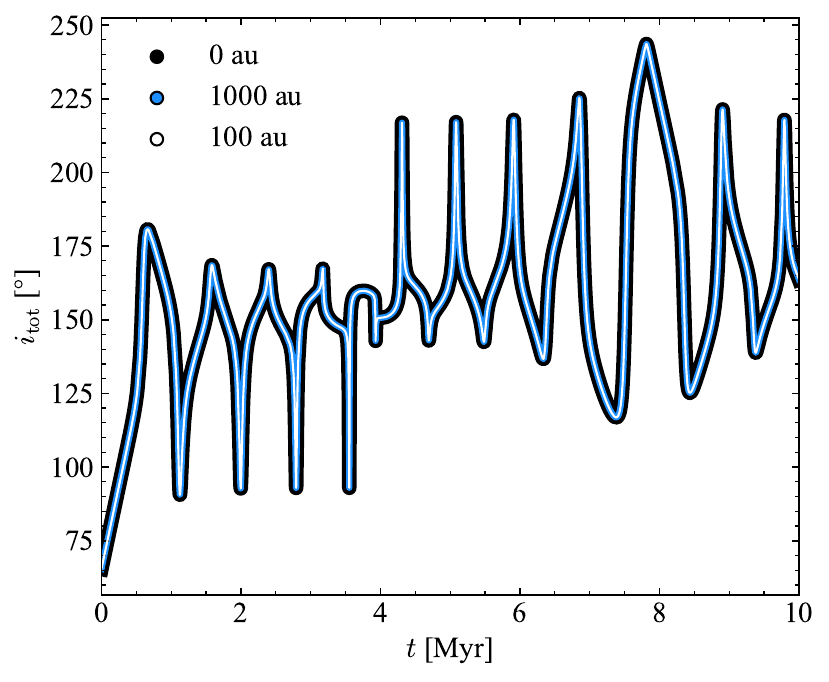}
            \includegraphics[width=\columnwidth]{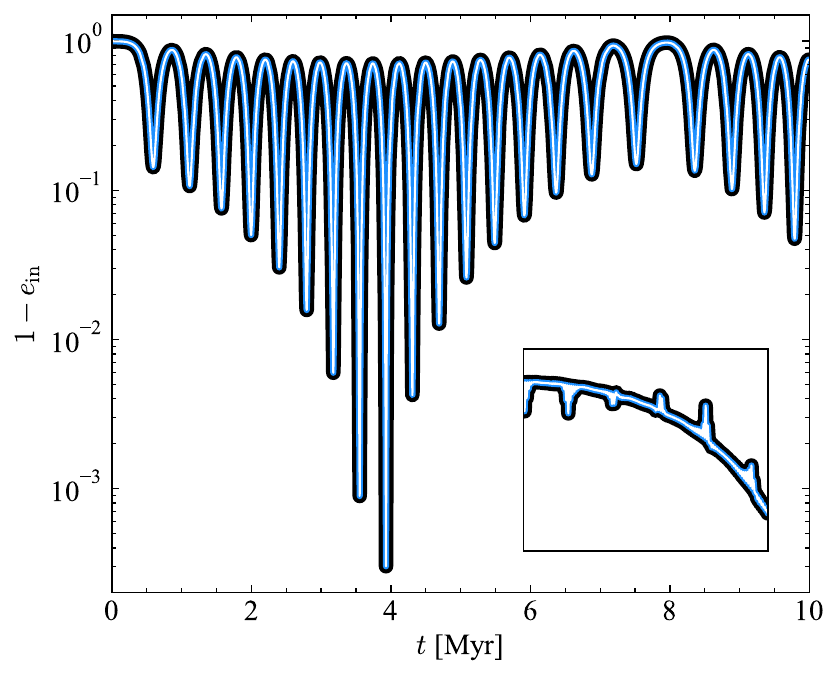}
            \caption{The total inclination of the triple system (top) and the inner binary's eccentricity (bottom) over time. The system comprises of an inner binary ($M_1=1$ M$_\odot$, $M_2=40$ M$_{\rm Jup}$) perturbed by an outer binary of mass $M=40$ M$_{\mathrm Jup}$. The inset shows the oscillation in $(1-e_{\rm in})$ for the first $0.12$ Myr. A cycle is found to be $0.821$ Myr. The $y$-axis corresponds to $\Delta (1-e)=3\times10^{-3}$, showing the extent which \texttt{Nemesis} can capture the ZLK cycles.}
            \label{Fig:ZLK_Ecc} 
        \end{figure} 
        Figure \ref{Fig:ZLK_Ecc} shows the oscillation of the total inclination ($i_{\rm tot}=i_{\rm out}+i_{\rm in}$) and the inner binaries' eccentricity over time for all three runs. Runs using a test particle as the planetary mass object yield identical results and in both cases, \texttt{Nemesis} resolves the ZLK effect over secular times at an accuracy indistinguishable from a direct $N$-body integrator (the inlet in the lower panel highlights the extent to which \texttt{Nemesis} replicates results from a direct $N$-body integration). A relative energy error of $\Delta E=3\times10^{-12}$ is found at the final step. 
        
        Theory estimates that the system considered here has a ZLK oscillation period of $0.913$ Myr \citep{2013ApJ...773..187N}. The estimate uses a double-averaged Hamiltonian, and as such is an approximation. Additionally, they note that when the inner binary reaches large eccentricities (as the case here), the approximation can change by `orders of magnitude'. Even so, the value is near the oscillation timescale seen here ($0.821$ Myr). Even with the consistent splitting of the child system and merging of parent particles, the procedure doesn't affect the dynamics, with a smooth evolution in the orbital parameters being observed. The result also highlights the validity of representing children with a CoM parent in the parent code.
        
        The capability to capture the ZLK effect is important for resolving secular timescales for dynamical systems. For instance, in planetary systems, the ZLK effect has been proposed to create hot Jupiters \citep{1999AJ....117..621H, 2017MNRAS.468.3000M} and has been considered for the evolution of asteroids \citep{2012ApJ...754L..36N, 2004Icar..172..372F}. Meanwhile, in the galactic nuclei, the ZLK effect will drive black hole-black hole mergers or hyper velocity stars \citep{2013ApJ...773..187N,  2018MNRAS.480.5160F, 2019ApJ...871...91Z, 2021A&A...652A..54A}. For an overview see \citet{2016ARA&A..54..441N}.
        
    \label{Sec:Nemesis}
    \subsection{\texttt{Nemesis} Scaling Relations}\label{Sec:Scaling}
        \subsubsection{Bridge Time Step}
\begin{figure}
            \centering
            \includegraphics[width=\columnwidth]{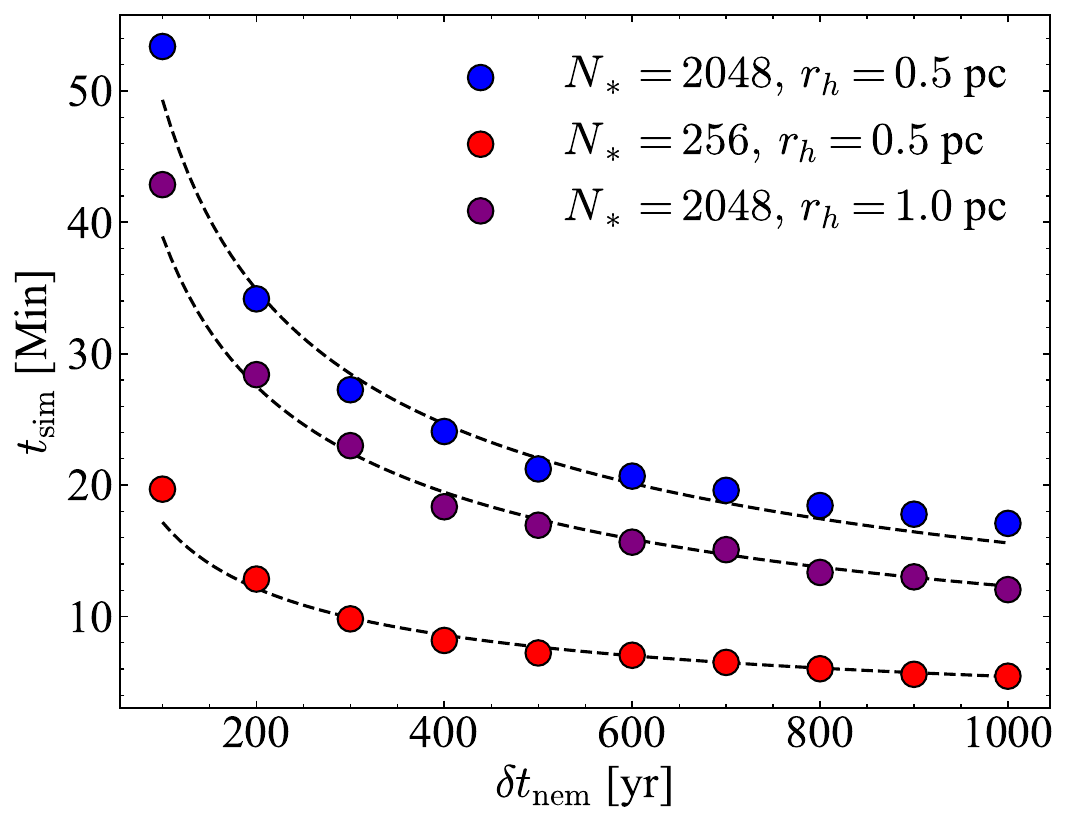}
            \includegraphics[width=\columnwidth]{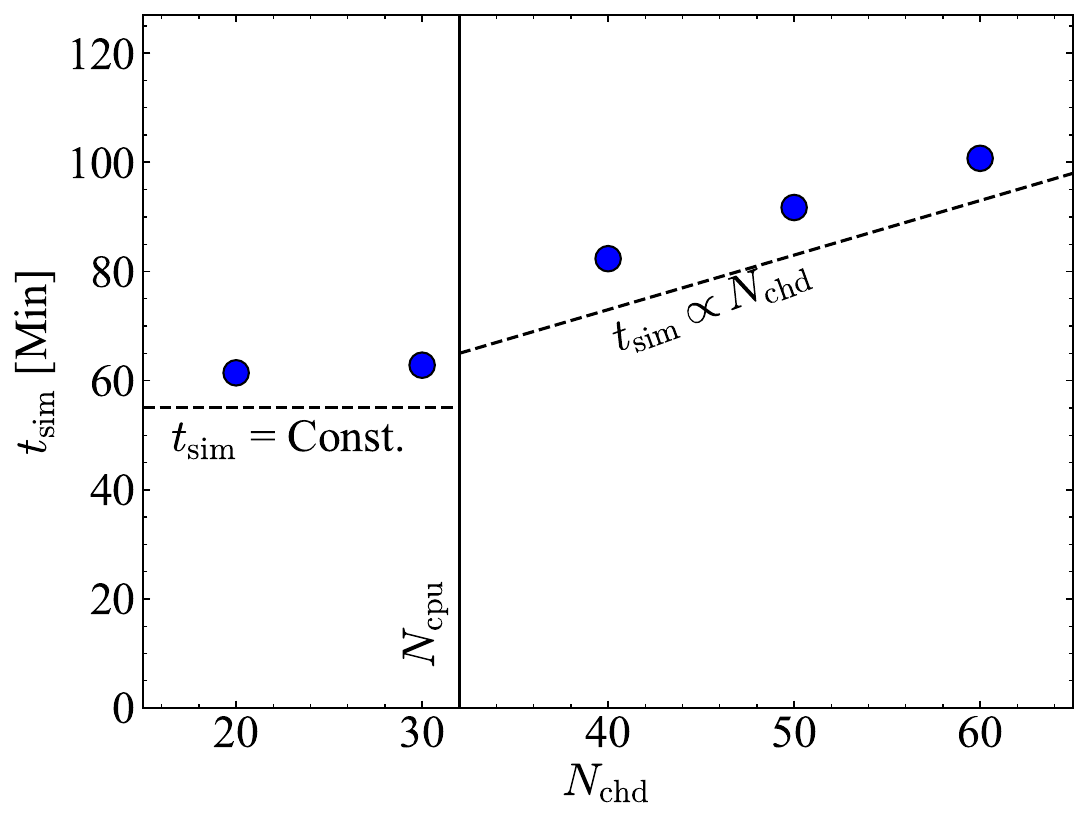}
            \caption{Top: \texttt{Nemesis} bridge time step $\delta t_{\rm nem}$ vs. computation time for an identical cluster containing 20  subsystems. Blue points consider a cluster with  density $\rho\sim10^{3}$ M$_\odot$ pc$^{-3}$, while red and purple points $\rho\sim10^{2}$ M$_\odot$ pc$^{-3}$. Dashed lines are curves described by the exponent $\alpha=-0.5$. Bottom: Number of children $N_{\rm chd}$ vs. time taken to simulate an identical cluster to $100$ kyr, $t_{\rm sim}$.}
            \label{Fig:scaling}
        \end{figure}

        The top panel of figure \ref{Fig:scaling} shows how \texttt{Nemesis} performs when changing the bridge time step, $\delta t_{\rm nem}$. As a reminder, this is the time when children and parents synchronise and communicate their respective gravitational potentials with one another.
        
        The relation $t_{\rm sim}\propto\delta t_{\rm nem}^{\alpha}$ is fitted with power-laws $\alpha=-0.52\pm0.02$ for the blue points, $\alpha=-0.58\pm0.02$ for the red points and $\alpha=-0.56\pm0.01$ for the purple points. The curves, however, all consider $\alpha=-0.5$ since these too provide reasonable fits. The slope depends on what dominates the calculation. For large $N_*$, the gravitational code starts to dominate run time, while for low $N_*$, algorithmic overhead will play a larger role. This is reflected in the slightly shallower slopes of the high-density, massive cluster runs, since algorithmic overhead now plays a smaller role.

        There is no optimum time step. The user should conduct test runs to find the optimal parameter. After all, a balancing act between computational resources saved by increasing $\delta t_{\rm nem}$ and loss of accuracy exists. Ideally, \texttt{Nemesis} would use an adaptive algorithm. For instance, as a cluster evolves and expands, long-range perturbations become increasingly weaker such that $\delta t_{\rm nem}$ can relax to larger values. As mentioned at the end of section \ref{Sec:DirectNbody}, although an adaptive time step scheme is not yet implemented, it is in the plans for future work.

        \subsubsection{Number of Children}
        The right panel of Figure \ref{Fig:scaling} shows how $N_{\rm chd}$ affects computation time. Here, $N_{\rm CPU}=32$ worker processes are executed using MPI.
        
        As expected, while $N_{\rm chd}< N_{\rm CPU}$, the run-time is approximately constant. However, when $N_{\rm chd}>N_{\rm CPU}$, the scaling becomes linear (the best fit giving a power-law $\alpha=0.978\pm0.002$). The linear trend represents a worst-case scenario, where all children require roughly the same computational time to integrate, and results from child codes hibernating until resources (linear in $N_{\rm CPU}$) become available. In theory, given the hardware of certain supercomputers, the scaling implies that simulating globular clusters where each star hosts a planetary system is now attainable. 
        
        As a final comparison between a direct $N$-body method and \texttt{Nemesis}, table \ref{Tab:CompTime} compares the time needed to simulate identical systems to $0.1$ Myr.
        
        \begin{table}
             \caption{Wall-clock time taken to simulate $N_{\rm chd}$ systems till $0.1$ Myr.}
            \label{Tab:CompTime} 
            \centering 
            \begin{tabular}{c c c c}
                \hline\hline
             $N_{\rm chd}$ & \texttt{Ph4} & \texttt{Nemesis} \\
                \hline \vspace{-0.75em}\\
               $10$ & $251$ mins    & $64$  mins   \\
               $20$ & $716$ mins    & $61$  mins   \\
               $60$ & $2059$ mins   & $101$ mins   \\\hline
            \end{tabular}
            \tablefoot{Col. 1: The number of children. Col. 2: Results using the direct $N$-body algorithm. Col. 3: Results using \texttt{Nemesis}.}
        \end{table}

        \subsection{Parent Radius}
        Although not directly tested, the choice of $R_{\rm par}$ also affects both performance and reliability. If $R_{\rm par}$ is too large, children are more likely to include multiple particles of comparable mass, which reduces the accuracy of the CoM approximation. Decreasing $\delta t_{\rm nem}$ can mitigate errors, but at the cost of increased computation time (figure \ref{Fig:scaling}) and without fully correcting the underlying issue. 
        
        Since the strength of \texttt{Nemesis} lies in decoupling small-scale systems from the large-scale environment, it is generally preferable to keep $R_{\rm par}$ small, the value of which should be tested for optimal performance. There are several reasons behind choosing small $R_{\rm par}$ (at the cost of relatively larger $\delta t_{\rm nem}$); 
        \begin{enumerate}
            \item \textbf{Centre-of-Mass Approximation}: As mentioned, a smaller $R_{\rm par}$ makes the CoM approximation more successful since it's less likely any one system has two massive particles with a similar mass-ratio, $q\equiv m_i/m_j$. In the situation that a system has the mass ratio between its two most massive particles $q\xrightarrow[]{}1$, then the omission of correction kicks between particles of two distinct child systems will start to break down. Similarly, too large an $R_{\rm par}$ means that wide orbits play a large role in shifting the CoM, resulting in a similar reduction in accuracy.
            \item \textbf{Avoiding Bottlenecks from Tight Orbits}: The problem of directly integrating small-scale systems in large-scale environments (i.e., planetary systems in clusters) is the need to suppress the drift in energy errors to allow capture of secular effects. This can be done with symplectic codes. However, with their shared time step scheme, the tightest orbits induce a bottleneck in the integration. With \texttt{Nemesis}, as long as $R_{\rm par}$ exceeds a typical system's tightest orbit, tight orbits are always decoupled from the global integration, removing the bottleneck for the parent integration while allowing for accurate integration of the internal dynamics. This scheme is indifferent to the extent $R_{\rm par}$ exceeds the scale of the tightest orbit.
            \item \textbf{Parent Particle Merging Events}: \texttt{AMUSE}'s collision handler internally checks at every time step whether two particles collide. Since cluster codes utilise an adaptive time step, scaling with the particle's free-fall time (see equation \ref{Eqn:Internaldt}), unless exceedingly small, all collision events within the parent are detected.
        \end{enumerate}
        Throughout testing, it was found that a radius coefficient $A=100$ sufficed (see equation \ref{Eqn:Rpar}). This was tested for environments ranging in densities between $\rho\sim(10^{2}-10^{4})\, \mathrm{M}_\odot\, \mathrm{pc}^{-3}$ but may change depending on the initial cluster parameters, i.e., initial virality or compactness of systems. 
        
        That said, in section \ref{Sec:DirectNbody} it was seen that particles with semi-major axis $a\sim R_{\rm par}$ are poorly resolved. As such, given the points listed above, it could be worthwhile to use a lower value. While this doesn't fix the underlying problem as child particles will still exist with $a\sim R_{\rm par}$, particles with smaller semi-major axis are better protected from the external environment and so deviations should be less substantial.
        
        A conservative $R_{\rm par}$ implies that wide systems, i.e., wide planets or Oort Cloud-like environments \citep{1932PAAAS..67..169O, 1950BAN....11...91O}, fall back to the parent code. While this does not resolve the local dynamics as successfully, the long secular timescales of wide orbits partly compensate for the loss in accuracy. Moreover, such systems are inherently fragile, and external tides or encounters with field particles are likely to strip them before secular effects become significant. There is a delicate balance between $\delta t_{\rm nem}$ and $R_{\rm par}$, and it remains for future work to find a logical and adaptable way to balance the two for a given system, keeping in mind computation time and accuracy of results.
        
    \section{Conclusions}
        The multi-scale, multi-physics nature of astronomy complicates computational astrophysics. This paper formally introduces an updated and optimised version of the hybrid code \texttt{Nemesis}, which has previously been used to simulate planetary systems in an Orion‐like cluster \citep{2019A&A...624A.120V}.

        After describing the motivation and theory behind the scheme in section \ref{Sec:Theory}, a suite of simulations are run to test its performance. The first test considers asteroids within a star cluster's evolution. When comparing results with a direct $N$-body code, the final orbital parameters of asteroids between algorithms were found to be statistically indistinguishable. When tracking the energy error over longer periods of time, \texttt{Nemesis} successfully suppresses energy drift when using symplectic codes. This feature is essential for simulating systems over secular timescales. The success in modelling secular effects is highlighted by the second set of tests, with its ability to reproduce the ZLK oscillations in hierarchical triples.
        
        Finally, two scaling tests are conducted. The first considers how wall-clock time scales with the bridge time step, $\delta t_{\rm nem}$. While smaller $\delta t_{\rm nem}$ lowers integration error, it increases algorithmic overhead, with the wall-clock time scaling as $t_{\rm sim}\propto1/\sqrt{\delta t_{\rm nem}}$. Users should therefore do trial runs before production runs to determine the optimal $\delta t_{\rm nem}$ since this may depend on the environment. Indeed, the virial state, density and/or binary fraction may all play a role. The second scaling test looked at how the number of subsystems present, $N_{\rm chd}$, influences the wall-clock time. Runtime remains constant while $N_{\rm chd}$ remains below the number of available CPUs. Beyond this point, computational cost scales at worst linearly. This behaviour emphasises the highly parallelised nature of \texttt{Nemesis} and its strength to integrate environments hosting numerous subsystems.
        
        Software is never finished, but even in its current form, \texttt{Nemesis} provides researchers with a practical way to tackle multi-scale and multi-physics problems. Given the flexibility of the algorithm and the \texttt{AMUSE} environment hosting numerous codes in the domain of hydrodynamics, gravity, radiative transport, chemistry and stellar evolution codes, creative solutions to complex problems can now be attainable.

    \begin{acknowledgements}
        We thank the referee for providing nice suggestions which improved the quality and clarity of this paper.
        
        Analysis was made using the open-source \texttt{Python} packages \texttt{NumPy} \citep{2020Natur.585..357H} and \texttt{Matplotlib} \citep{2007CSE.....9...90H}.
    \end{acknowledgements}

\bibliographystyle{aa}
\bibliography{references.bib}

\end{document}